\documentclass[AMA,STIX1COL]{WileyNJD-v2}
\usepackage{moreverb}
\usepackage{amsfonts}
\usepackage{latexsym,amsmath}
\usepackage{graphicx}

\newcommand\BibTeX{{\rmfamily B\kern-.05em \textsc{i\kern-.025em b}\kern-.08em
T\kern-.1667em\lower.7ex\hbox{E}\kern-.125emX}}

\articletype{Article Type}%

\received{<day> <Month>, <year>}
\revised{<day> <Month>, <year>}
\accepted{<day> <Month>, <year>}

\definecolor{lgray}{gray}{0.45}
\newcommand{\be}{\begin{equation}}
\newcommand{\ee}{\end{equation}}
\newcommand{\bea}{\begin{eqnarray}}
\newcommand{\eea}{\end{eqnarray}}

\begin{document}

\title{Nonlinear instability and solitons in a self-gravitating fluid}

\author[1]{G. N. Koutsokostas}

\author[2,3]{S. Sypsas}

\author[2,4]{O. Evnin}

\author[5]{T. P. Horikis}

\author[1]{D. J. Frantzeskakis}

\authormark{G. N. Koutsokostas \textsc{et al}}

\address[1]{\orgdiv{Department of Physics}, \orgname{National and Kapodistrian University of Athens}, \orgaddress{\state{Athens}, \country{Greece}}}

\address[2]{\orgdiv{High Energy Physics Research Unit, Faculty of Science}, \orgname{Chulalongkorn University}, \orgaddress{\state{Bangkok}, \country{Thailand}}}

\address[3]{\orgdiv{NARIT}, \orgname{Don Kaeo, Mae Rim}, \orgaddress{\state{Chiang Mai}, \country{Thailand}}}

\address[4]{\orgdiv{Theoretische Natuurkunde}, \orgname{Vrije Universiteit Brussel and International Solvay Institutes}, \orgaddress{\state{Brussels}, \country{Belgium}}}

\address[5]{\orgdiv{Department of Mathematics}, \orgname{University of Ioannina}, \orgaddress{\state{Ioannina}, \country{Greece}}}

\corres{*D. J. Frantzeskakis, Department of Physics, National and Kapodistrian University of Athens, Panepistimiopolis, Zografos, Athens 15784, Greece. \email{dfrantz@phys.uoa.gr}}

\abstract[Abstract]{

We study a spherical, self-gravitating fluid model, 
which finds applications in cosmic structure formation. 
We argue that since the system features nonlinearity and gravity-induced dispersion, 
the emergence of solitons becomes possible. We thus employ a multiscale expansion 
method to study, in the weakly nonlinear regime, the evolution of small-amplitude 
perturbations around the equilibrium state. This way, we derive a 
spherical nonlinear Schr{\"o}dinger (NLS) equation that governs the envelope of the 
perturbations. The effective NLS description allows us to predict a ``nonlinear 
instability'' (occurring in the nonlinear regime of the system), namely, the 
modulational instability which, in turn, may give rise to spherical soliton states. 
The latter feature a very slow (polynomial) curvature-induced decay in time. The 
soliton profiles may be used 
to describe the shape of dark matter halos at the rims of the galaxies.

}

\maketitle

\section{Introduction} \label{intro}

A set of strongly-interacting particles on scales much larger than their mean 
free-path, i.e., where the motion of each individual particle relative to its 
neighbors is negligible, can be described via conservation laws. Within this 
description, the collective behavior of the system will be governed by the 
equations of fluid mechanics.

One of the simplest possible cases in this context is an isentropic one-dimensional 
(1D) flow, described by the following system of partial differential equations (PDEs) 
---see, e.g., Refs.\cite{courfr,whitham}: the continuity equation (conservation 
of mass),
\begin{eqnarray}
\frac{\partial \rho}{\partial t} + \frac{\partial}{\partial x}(\rho u)=0,
\label{ce}
\end{eqnarray} 
where $\rho(x,t)$ is the mass density and $u(x,t)$ is the fluid velocity, 
as well as the Euler equation (conservation of momentum),
\begin{eqnarray}
\frac{\partial u}{\partial t} + u \frac{\partial u}{\partial x}
= -\frac{1}{\rho}\frac{\partial p}{\partial x},
\label{Ee}
\end{eqnarray} 
where $p(x,t)$ is the pressure. The latter is connected to the density
via an equation of state,
\begin{equation}
p=p(\rho),
\label{Es}
\end{equation} 
where it is assumed that $dp/d\rho=c_s^2(\rho)>0$, with $c_s$ being the (generally 
density-dependent) speed of sound. The nonlinear system~(\ref{ce})-(\ref{Es}) 
can be expressed in the following conservation form:
\begin{eqnarray}
&&\frac{\partial \rho}{\partial t} + \frac{\partial}{\partial x}(\rho u)=0, 
\label{ce2} \\
&&\frac{\partial}{\partial t}(\rho u) + \frac{\partial}{\partial x}
\left[\rho u^2 +p(\rho) \right]=0,
\label{Ee2}
\end{eqnarray}     
which is in fact identical to the compressible isentropic gas dynamical equations, and 
can be solved by the ingenious method introduced by Riemann \cite{riemann}. Indeed,
the hyperbolic Eqs.~(\ref{ce2})-(\ref{Ee2}) can be diagonalized by the Riemann 
invariants: 
\begin{eqnarray}
r_1=u-\int^{\rho} \frac{c_s(\rho')}{\rho'}d\rho', \quad  
r_2=u+\int^{\rho} \frac{c_s(\rho')}{\rho'}d\rho', 
\label{r12}
\end{eqnarray} 
with characteristic velocities 
\begin{equation}
v_1 = u-c_s(\rho), \quad v_2 = u+c_s(\rho), 
\label{charvs}
\end{equation}
so that
\begin{equation}
\frac{\partial r_j}{\partial t} + v_j \frac{\partial r_j}{\partial x}=0, \quad j=1,2.
\label{RIe}
\end{equation} 
Due to the monotonicity 
\footnote{Generally, for a polytropic gas, $p(\rho)=A\rho^\gamma$
(where $\gamma$ is the polytropic index and $A={\rm const.}$), one finds 
$g(\rho)=[2/(\gamma-1)]c_s(\rho)$, with $c_s(\rho)$ being a monotone function.} 
of $g(\rho)=\int^{\rho} \left(c_s(\rho')/\rho'\right)d\rho'$,
Eqs.~(\ref{r12}) can be inverted leading to:
\begin{equation}
u=\frac{1}{2}(r_1+r_2), \quad \rho=g^{-1}\left(\frac{1}{2}(r_2-r_1) \right).
\end{equation}
The above purely nonlinear hyperbolic system is a prototypical setting where smooth 
initial data may develop singularities in finite time, i.e., shock waves. In this 
regard, the existence of Riemann invariants allows for estimates of the breaking times 
at which these may occur\cite{Lax}.  

Nevertheless, apart from the intrinsic nonlinearity which may give rise to purely 
nonlinear waves in the form of shocks, the fluid model under consideration may also 
display the important phenomenon of dispersion if gravity is included. Indeed, let us 
follow Ref.\cite{kates,ononakata}, and introduce a 1D self-gravitating gas, such that 
the Euler equation is modified as
\begin{eqnarray}
\frac{\partial u}{\partial t} + u \frac{\partial u}{\partial x}
= -\frac{1}{\rho}\frac{\partial p}{\partial x}-\frac{\partial \Phi}{\partial x},
\label{Eeon}
\end{eqnarray} 
where $\Phi$ is the gravitational potential obeying the Poisson equation: 
\begin{equation}
\frac{\partial^2 \Phi}{\partial x^2} = 4\pi G \left( \rho-\rho_0 \right).
\label{1dP}
\end{equation}
Here, $G$ is Newton's gravitational constant and $\rho_0$ is an equilibrium density. 
Since we are interested in fluctuations propagating on top of the equilibrium 
background density, let us define the (linear) overdensity as 
$\rho_1 \equiv (\rho-\rho_0)/\rho_0$. Then, the pertinent PDEs describing this self-gravitating 
gas, namely Eqs.~(\ref{ce}), (\ref{Eeon}), (\ref{1dP}), together with the 
state-equation~(\ref{Es}), when linearized around the equilibrium state, admit plane 
wave solutions $\rho_1 \propto \exp[i(kx-\omega t)]$, with the 
frequency $\omega$ and the wavenumber $k$ satisfying the dispersion relation
\begin{equation}
\omega^2(k) = c_s^2(\rho_0) \left(k^2 -k_J^2\right),
\label{JIdr}
\end{equation}
where $k_J=(4\pi G \rho_0)^{1/2}/c_s(\rho_0)$ is the so-called Jeans wavenumber.
Since $\partial^2 \omega(k)/\partial k^2 \neq 0$, the nonlinear system is also 
dispersive\cite{whitham}, a fact that suggests the 
possibility of soliton formation. 
%

Apart from highlighting the presence of gravity-induced dispersion, the 
relation~(\ref{JIdr}) also suggests the occurrence of the (linear) Jeans instability
\cite{Jeans}: perturbations of wavenumbers $|k|<k_J$ are unstable and grow 
exponentially in time. Thus, at scales larger than the Jeans length, where 
gravity dominates the internal pressure, overdensities collapse under their 
gravitational pull leading to the emergence of structures.


Recently, there has been a substantial increase of interest in self-gravitating 
fluid models, as they generically appear in studies in astrophysics and plasmas 
\cite{chavanis2}, as well as in cosmology, and particularly in the context of 
cold dark matter ---see, e.g., Refs.~\cite{witten,harko,chavanis1,mocz}. In such 
settings, obviously, pertinent self-gravitating fluid systems are 
considered in higher-dimensions (and in many cases in a spherical geometry). 
However, once the dynamics is considered in the fully nonlinear regime, due to 
their complicated character, multidimensional self-gravitating fluid models are 
usually studied by means of numerical simulations 
---see, e.g., Refs.~\cite{mocz,natphys}.

Here, motivated by these works, our scope is to study analytically a radial 
three-dimensional (3D) fluid model [see Eqs.~(\ref{1o})-(\ref{3o}) and (\ref{P-r}) 
below]. Upon employing a multiscale expansion method we are able to approximate 
the system, under certain conditions, with a much more tractable equation. In 
particular, we show that, in the weakly nonlinear regime, small-amplitude excitations 
around the equilibrium state of the fluid are described, at slow length/time scales, by an effective {\it spherical} nonlinear Schr{\"o}dinger (NLS) equation of the form:
\begin{equation}
i \frac{\partial A}{\partial \tilde{t}} 
+ i\frac{A}{\tilde{t}}
+\frac{1}{2}\frac{\partial^2 \omega(k)}{\partial k^2} 
\frac{\partial^2 A}{\partial \tilde{r}^2}+\gamma(k) |A|^2 A=0,
\label{NLS0}
\end{equation} 
where $A$ is the envelope of the perturbations, $\tilde{t}$ and $\tilde{r}$ denote 
proper slow time and slow radial variable, $\omega(k)$ is given by Eq.~(\ref{JIdr}), 
and $\gamma(k)$ is the nonlinearity coefficient, which we specify below. Notice that 
the spherical NLS~(\ref{NLS0}), which has been used to describe radial solitons in 
plasmas~\cite{shukla1,xue1} and fluids~\cite{huang}, is actually a traditional NLS 
equation, which also incorporates the curvature term $iA/\tilde{t}$ (in fact, as we 
will see, $r^{-1} \propto \tilde{t}^{-1}$) stemming from the consideration of 
the radial $3$D geometry. This term suggests that, although the initial system is 
conservative, the underlying wave structures feature a curvature-induced decay (see 
details in Sec.~5 below). This effect is reminiscent to the situation occurring, e.g., 
in the usual radially symmetric 2nd-order wave equation, whose spherical 
(D'Alembert) wave solutions feature a curvature-induced decay $\propto 1/r$.  

With the effective spherical NLS model at hand, we are able to deduce the following. 
First, we investigate a certain ``nonlinear instability'', i.e., one occurring in the 
nonlinear regime of the system, the so-called modulational instability (MI) 
\cite{ZAKHAROV2009540} that can manifest itself in addition to the Jeans instability. 
This way, we find that the weak, spatially uniform and time-varying solution of the 
NLS is ---under certain conditions--- subject to MI, for wavenumbers outside the Jeans 
instability band, namely for $k_J <|k| < \alpha k_J$, with $\alpha \approx 1.34$. 
As mentioned, this is an example of nonlinearity (i.e., the nonlinearity-induced MI) 
competing with the gravity-induced dispersion, giving rise to the formation of 
solitons. The latter may appear as regular sech-shaped solitons\cite{zakharov} or 
rogue waves\cite{onorato_report,tov} and they are characterized by an exponential or 
an algebraic decay in space, respectively, while both experience a slow curvature-
induced polynomial decay in time. 


Physically, our results could be relevant for cosmic structure formation. The system~(\ref{ce}), (\ref{Eeon}), 
(\ref{1dP}), extended to three dimensions is a fair description of the universe during its (dark) matter-dominated era. Small density fluctuations along the initially homogeneous background, provide the seeds that grow to form local dark-matter overdensities, objects which then act as gravitational wells that attract (luminous) baryonic matter, eventually forming the galaxies we observe. The radial distribution of dark matter (DM) in these galactic halos is inferred by the rotational motion of their baryonic component\cite{rubin2004} and seems to hint upon a layered structure: a central, smooth, high-density core followed by a decaying tail\cite{de_Blok_2010}. In pure dark matter models, the presence of a core can be attributed to the Jeans instability supported by pressure\cite{Schive:2014dra}. In this context, the main claim of the present work is the appearance of MI-induced solitonic structures at scales larger than the Jeans length; for a cored DM halo this might appear as a slightly overdense outer shell. 

The paper is organized as follows. First, in Sec.~2 we introduce the fluid model, 
and study the linear regime, where the Jeans instability occurs. In Sec.~3, using a 
multiscale expansion method, we derive the effective spherical NLS equation. In 
Sec.~4, we analyze the nonlinear instability (MI) that is predicted in the framework 
of the NLS, while in Sec.~5 we present spherical soliton 
solutions that can be formed due to MI. Finally, in Sec.~6 we summarize our 
findings and discuss possibilities for future work.

\section{The model and Jeans instability}

We consider the dynamical equations for an ideal, multi-dimensional fluid, in the 
context of non-relativistic Einstein gravity. These include the continuity equation: 
\begin{equation}
\frac{\partial \rho}{\partial t} + \mathbf{\nabla} \cdot (\rho \mathbf{u})=0,
\label{1o}
\end{equation}  
where $\rho(\mathbf{r},t)$ and $\mathbf{u}(\mathbf{r},t)$ denote the mass density 
and velocity, respectively, as well as the Euler equation:
\begin{equation}
\frac{\partial \mathbf{u}}{\partial t} + \left(\mathbf{u} \cdot \mathbf{\nabla}\right) \mathbf{u} = 
-\frac{1}{\rho} \mathbf{\nabla} p - \mathbf{\nabla} \Phi.
\label{2o} 
\end{equation}
Here, the gravitational potential $\Phi(\mathbf{r},t)$ is coupled to the 
mass density via the Poisson equation,
\begin{equation}
\Delta \Phi = 4\pi G \left( \rho-\rho_0 \right),
\label{3o}
\end{equation}  
where, as above, $G$ is Newton's gravitational constant, $\rho_0$ is the equilibrium density and $\Delta=\nabla^2$. 
Finally, the pressure $p$ in Eq.~(\ref{2}) is related to the mass density $\rho$ 
by an equation of state, $p=p(\rho,T)$, with $T$ being the temperature. We consider 
the case of constant temperature and adopt a phenomenological model, where 
the pressure and the density obey the following equation:
%
%
\begin{equation}
p(\rho)= c_s^2 \rho,
\label{4o}
\end{equation}
with $c_s$ being the speed of isothermal sound, which is assumed to be constant in 
what follows. This assumption corresponds to Boyle's law for ideal gases, and is relevant to the considered dark matter problem for the 
following reasons. First, we note that the state equation~(\ref{4o}) corresponds to 
the case of a spherical polytropic gas, and suggests that $c_s^2=kT/m$, where $k$ 
is the Boltzmann's constant, $T$ is the temperature of the dark matter (assumed to be 
constant and very low), and $m$ is the mass of the dark matter particles. 
In the standard cosmological model, 
known as Lambda-cold-dark-matter ($\Lambda$CDM), dark matter is assumed to consist of 
cold, non-interacting particles, the mass of which is currently unknown. In a fluid 
language, ``cold'' translates to a vanishing sound speed and, consequently, the Jeans 
length vanishes too. However, a 
speed of sound can be attributed\cite{Armendariz-Picon:2013jej} due to (isotropic) velocity dispersion: $c_s=\sqrt{\langle v_i^2\rangle}$, which suggests that 
$c_s=$~const. On the other hand, another possibility with a non-vanishing Jeans length is the so-called fuzzy dark matter\cite{Hu:2000ke}, which considers the dynamics of a nonrelativistic, ultra-light scalar field coupled to gravity. In fact, our qualitative results depend only on the existence of the Jeans length and are realization independent.
%


It is convenient for our analysis to express the model~(\ref{1o})-(\ref{3o}) 
and (\ref{4o}) in dimensionless form. This can be done upon measuring the density 
$\rho$ and pressure $p$ in units of their equilibrium values $\rho_0$ and 
$p_0=p(\rho_0)$, and normalize the velocity $\mathbf{u}$, the 
gravitational potential $\Phi$, the space variable $\mathbf{r}$ and time $t$ with 
respect to 
$c_s$, $c_s^2$, $(p_0/4\pi G \rho_0^2)^{1/2}$ and $(4\pi G \rho_0)^{-1/2}$, 
respectively. In what follows, we set the sound speed to 
$c_s = 10^{-6}c_0 =3\times10^2~{\rm m/s}$, with $c_0$ the speed of light; such a value 
is compatible with observations\cite{Muller:2004yb}. Then, the above units take the 
following values: the equilibrium density is 
$\rho_0 \approx 2.5 \times 10^{-27}~{\rm kg/m^3}$, while 
the characteristic length and time scales are 
$7.16\times 10^{22}~{\rm m} \approx 7~{\rm kpc}$ and 
$6.9 \times 10^{17}~{\rm s} \approx 2.2 \times 10^{10}~{\rm Julian~years}$ 
[the gravitational constant is $G=6.67 \times 10^{-11}~{\rm m^3/(kg \cdot s^2)}]$. 
As we will show later, the structures we will be focusing on have characteristic sizes 
a few times larger than the Jeans radius, placing them at the edge of typical 
galactic halos. Finally, as per common practice, since the gravitationally 
bound objects we will be discussing decouple from the expanding background, in what 
follows we will neglect the cosmological expansion and treat spacetime as static.  


Under this assumption and in these units, the considered fluid model is 
written in the following dimensionless form:

%
\begin{eqnarray}
&&\frac{\partial \rho}{\partial t} + \mathbf{\nabla} \cdot (\rho \mathbf{u})=0,
\label{1} \\
&&\frac{\partial \mathbf{u}}{\partial t} + (\mathbf{u} \cdot \mathbf{\nabla}) \mathbf{u} = 
-\frac{1}{\rho} \mathbf{\nabla} p - \mathbf{\nabla} \Phi,
\label{2} \\
&&\Delta \Phi = \rho-1,
\label{3} \\
&&p(\rho)=\rho.
\label{4}
\end{eqnarray}
A simple nontrivial solution of the above system is
\begin{equation}
\rho=1, \quad p=1, \quad \mathbf{u}=0, \quad \Phi=0,
\end{equation}
which corresponds to a stationary background solution. 

To study the stability of the background, we introduce the perturbation ansatz
\begin{equation}
\rho=1+\epsilon \rho_1, \quad p = 1 +\epsilon p_1, \quad 
\mathbf{u}= \epsilon \mathbf{u}_1, \quad \Phi= \epsilon \Phi_1,
\label{3aa}
\end{equation}
where $0 <\epsilon \ll 1$ is a formal small parameter denoting the amplitude of 
fluctuations with respect to the background, while $\rho_1$, $p_1$, $\mathbf{u}_1$, 
and $\Phi_1$ are arbitrary perturbations of order ${\cal O}(1)$. Substituting 
Eqs.~(\ref{3aa}) into Eqs.~(\ref{1})-(\ref{4}) we obtain at ${\cal O}(\epsilon)$ the 
following linear system:
\begin{eqnarray}
&&\frac{\partial \rho_1}{\partial t} + \mathbf{\nabla} \cdot \mathbf{u}_1=0,
\label{4a}\\
&&\frac{\partial \mathbf{u}_1}{\partial t} = -\mathbf{\nabla} p_1 - \mathbf{\nabla} \Phi_1,
\label{4b}\\
&&\Delta \Phi_1= \rho_1,
\label{4c} \\
&&p_1= \rho_1,
\end{eqnarray}
which can be reduced to a single equation for $\rho_1$, namely:
\begin{equation}
\frac{\partial^2 \rho_1}{\partial t^2} -\Delta \rho_1 -\rho_1=0.
\label{leq}
\end{equation}
Equation~({\ref{leq}) admits plane wave solutions, namely  
$\rho_1 \propto \exp[i(\mathbf{k} \cdot \mathbf{r}-\omega t)] +{\rm c.c.}$  
(where c.c. denotes complex conjugate),
with the perturbation's frequency 
$\omega$ and the amplitude $k=|\mathbf{k}|$ of the wavevector $\mathbf{k}$ 
obeying the dispersion relation:
\begin{equation}
\omega^2(k)=k^2-k_{J}^2,  
\label{ldrj}
\end{equation}  
where the parameter $k_{J}$ is the Jeans wavenumber and takes the value $k_{J}=1$ 
in our units (recall that, in physical units, the Jeans wavenumber reads 
$k_J=(4\pi G \rho_0)^{1/2}/c_s(\rho_0)$, as mentioned in the Introduction).   
Importantly, the dispersion relation~(\ref{ldrj}) implies that for perturbation 
wavenumbers less than $k_J$, i.e., for $|k|<k_{J}$, the frequency $\omega(k)$ becomes 
purely imaginary and, hence, the perturbation behaves as 
$\exp[{\rm Im}\,\omega(k) t] \exp(i \mathbf{k}\cdot \mathbf{r})$ 
(the other mode decays exponentially in time), featuring a growth rate:
\begin{equation}
{\rm Im}\,\omega(k) =\sqrt{1-k^2}.
\label{Jgr}
\end{equation} 
In such a case, the system is unstable, and the relevant instability is the so-called 
Jeans instability. Notice that the growth rate becomes maximum for $k = 0$ (i.e., for 
infinite wavelengths) and its value is ${\rm Im}\,\omega(0)=1$. 

\begin{figure}[t]
\centerline{\includegraphics[scale=0.4]{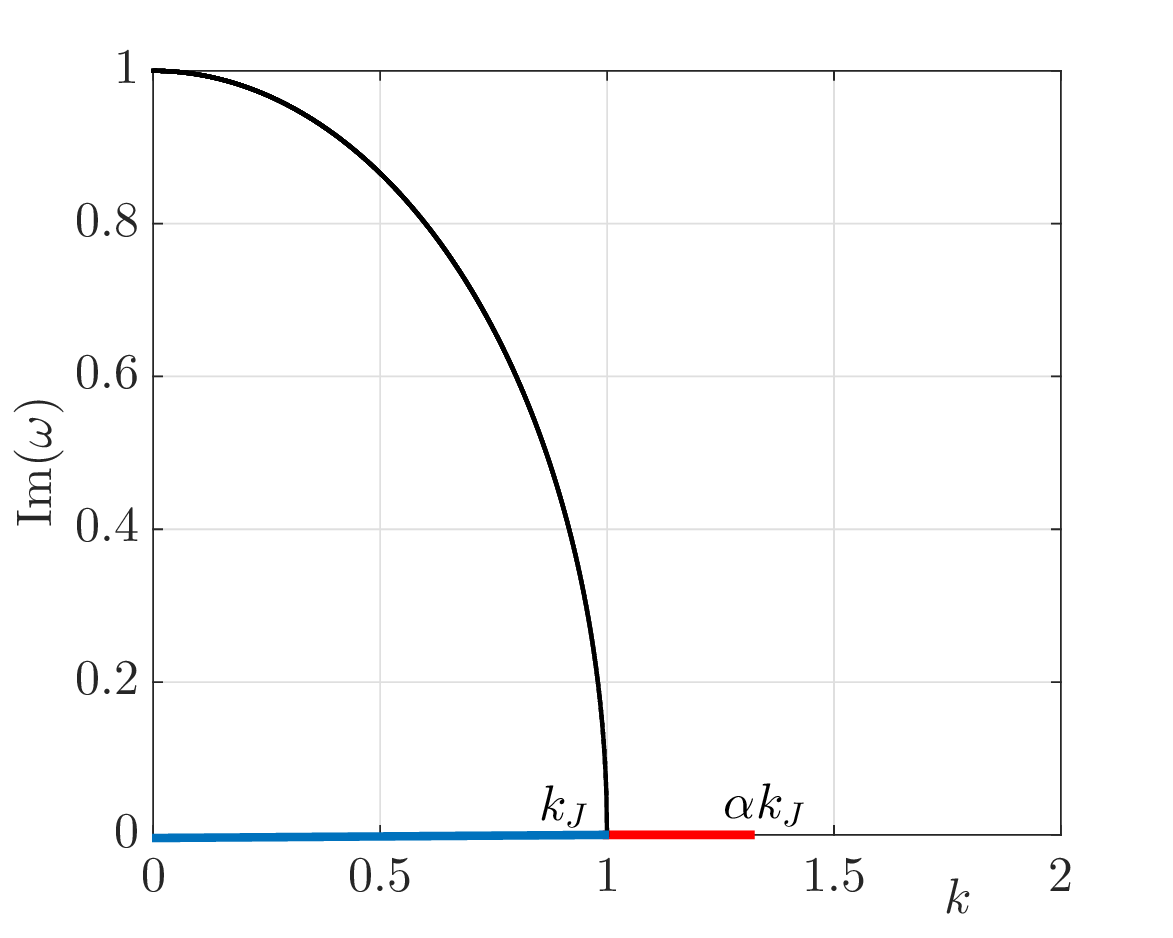}}
\caption{The Jeans instability growth rate ${\rm Im}\,\omega(k)$ as a function of the 
wavenumber $k>0$. According to the linear theory, the 
instability band is defined by the interval $k<k_{\rm J}=1$, highlighted with the blue segment. 
The red segment depicts the extension of the instability band once nonlinearity is taken into 
account ---see discussion in Sec.~\ref{nMI}.}
\label{imoF}
\end{figure}

The growth rate, ${\rm Im}\,\omega(k)$, as a function of $k$ is shown in Fig.~\ref{imoF}. Importantly, the growth rate drops to zero for $k=k_{J}=1$, implying that the Jeans instability 
ceases to exist for $|k| \geq 1$ (indeed, as seen from Eq.~(\ref{ldrj}), in this case 
$\omega(k) \in \mathbb{R} ~\forall k$). Nevertheless, as we will show in the next section, 
an analysis of the (weakly) nonlinear regime of the problem reveals that a nonlinear 
instability may occur for $k\gtrsim 1$. The relevant extension of the instability band 
beyond $k_{J}$ is depicted as a red horizontal line in Fig.~\ref{imoF} (see details below).

\section{Derivation of the effective spherical NLS equation}

The above discussion on the Jeans instability actually stems from the consideration 
of the linearized version of the self-gravitating fluid model. Our scope in this 
section is to show that it is also possible to study analytically the full, nonlinear 
problem. The latter, apart from being nonlinear, is also dispersive, since the 
dispersion relation~(\ref{ldrj}) implies that 
$\partial^2 \omega(k)/\partial k^2 \neq 0$. It is then our intention to employ  
the method of multiple scales\cite{jef} to asymptotically reduce the original 
self-gravitating fluid model to a simpler nonlinear evolution equation, namely, a 
spherical NLS equation, that admits soliton solutions.  

We start by noticing that, in the case of a spherical system under consideration, it 
is natural to seek for solutions of Eqs.~(\ref{1})-(\ref{4}) with angular symmetry; in such a case, the Laplacian in Eq.~(\ref{3}) may be expressed as
\begin{equation}
\Delta = \frac{\partial^2}{\partial r^2}+\frac{2}{r}\frac{\partial}{\partial r},
\label{P-r}
\end{equation}
%
while the fluid velocity reduces to $\mathbf{u}=(u,0,0)$. Notice that in this work 
we focus on the simplest nontrivial setup: a spherically symmetric overdensity. 
Here, $r=0$ represents the center of the object that will finally be formed. In a 
realistic scenario, there will be a collection of such objects and, hence, there, the 
spherical symmetry assumption breaks down. Strictly speaking, we are going to solve 
the problem on a ball of radius $r_0$ which is much larger than the radius of the 
object in question and much smaller than the distance between two objects. 

In the wavenumber regime $k>k_J=1$, where the system 
is still not prone to the Jeans instability, we seek solutions of the 
system~(\ref{1})-(\ref{3}) in the form of the following asymptotic expansions:
\begin{eqnarray}
\rho = 1 +\sum_{j=1}^{\infty} \epsilon^j \rho_{j}(r,t,R,T_2), \quad
u= \sum_{j=1}^{\infty} \epsilon^j u_{j}(r,t,R,T_2), \quad
\Phi = \sum_{j=1}^{\infty} \epsilon^j \Phi_{j}(r,t,R,T_2).
\label{6}
\end{eqnarray}
Here, the unknown functions, $\rho_j$, $u_j$ and 
$\Phi_j$, depend on the fast variables 
$r$ and $t$, as well as the slow variables $R_1$, $T_1$, and $T_2$, 
which are defined as follows:
\begin{equation}
R=R_1- V T_1, \quad R_1=\epsilon (r-r_0), \quad T_1=\epsilon t, 
\quad T_2=\epsilon^2 t,
\label{var}
\end{equation}
where $r_0$ is the radius of the initial configuration (e.g., 
the initial radius of a spherical soliton ---see below), while the unknown 
velocity $V$ in the definition of $R$ will be determined in a self-
consistent manner. Notice that $R=R_1-VT_1 = \epsilon(r-Vt-r_0)$ is 
actually a traveling wave coordinate associated with the expansion ($V>0$) or contraction ($V<0$) of the sought spherical structure. 
As such, it describes the increase or decrease of the spherical 
structure's radius with time. The domain of the variable $R$ is 
$-\epsilon(r_0+Vt)<R<+\infty$ (since $r>0$), and hence it can take both positive 
and negative values, for radii outside or inside the sphere of radius $r_0+Vt$.  
Furthermore, the above unknown fields 
are assumed to satisfy vanishing boundary conditions as $r\rightarrow \infty$. 

It is also relevant to make the following remarks. First, since the system features 
quadratic nonlinearities, the above slow variables are the only ones that appear at 
the third order of approximation [i.e., at ${\cal O}(\epsilon^3)$], where the first nonlinear 
equation emerges ---as we will see this will be the NLS equation. 
In addition, the choice of the slow traveling wave coordinate $R$ is made 
to facilitate the calculations at ${\cal O}(\epsilon^2)$ and ${\cal O}(\epsilon^3)$ ---see below.  
Furthermore, it is noticed that Eqs.~(\ref{var}) suggest that 
$1/r \sim \epsilon^2/(V T_2)$, which means that curvature effects will become 
important at ${\cal O}\!\left(\epsilon^3\right)$. Finally, it should be mentioned that  
our analysis is similar to the one in the $(1+1)$-dimensional setting 
\cite{ononakata}; nevertheless, in that work, the reductive perturbation method
---instead of the multiple scales technique [see Eqs.~(\ref{var})]--- is used, 
while the important, to our spherical setting, above mentioned curvature term 
was not taken into regard. As we will see, the inclusion of this term will lead us 
to a spherical NLS equation, rather the traditional one appearing in the $1$D setting.  

Substituting the expansions~(\ref{6}) into Eqs.~(\ref{1})-(\ref{3})
we obtain, at each order in $\epsilon$, the following system:
\begin{eqnarray}
&&\frac{\partial \rho_j}{\partial t} + \frac{\partial u_j}{\partial r} =f_j,
\label{7a}\\
&&\frac{\partial u_j}{\partial t} + \frac{\partial \rho_j}{\partial r}
+ \frac{\partial \Phi_j}{\partial r}  = g_j,
\label{7b}\\
&&\frac{\partial^2 \Phi_j}{\partial r^2} - \rho_j=h_j,
\label{7c}
\end{eqnarray}
where $j=1,2,\dots,$ denotes the order of the perturbation scheme, while the inhomogeneous
parts $f_j$, $g_j$ and $h_j$ will be defined below. Similarly to the analysis of the linearized 
version of the model, it is possible to derive from Eqs.~(\ref{7a})-(\ref{7c}) a single equation 
for the fields $\rho_j$, namely:
\begin{equation}
\frac{\partial^2 \rho_j}{\partial t^2 }- \frac{\partial^2 \rho_j}{\partial r^2 } -\rho_j =
\frac{\partial f_j}{\partial t} - \frac{\partial g_j}{\partial r} + h_{j}.
\label{eqrho} 
\end{equation}
\\
Below we focus on the first three orders of approximation, for which the above system
leads to the following results.

First, at $\mathcal{O}(\epsilon)$ (i.e., for $j=1$), the system (\ref{7a})-(\ref{7c}) is 
linear and homogeneous, namely, $f_1=g_1=h_1=0$; accordingly, Eq.~(\ref{eqrho}) (for $j=1$) 
becomes identical to the 1D version of Eq.~(\ref{leq}), with $\Delta =\partial_r^2$. 
Thus, a solution of Eq.~(\ref{eqrho}) for $j=1$ is of the form   
\be 
\rho_1 \sim A\exp(i\theta)\,,\quad \theta =kr-\omega t,
\ee
with $\omega$ and $k$ obeying the Jeans dispersion relation~(\ref{ldrj}) and $A$ 
being an unknown complex envelope function, which depends on the slow variables $R$ and $T_2$. 
Notice that for the above solution, we have considered the case of right-going waves, 
with $\omega=\sqrt{k^2-1}$.
In addition, it can readily be seen that the remaining fields $u_1$ and $\Phi_1$ take 
a similar form, and thus a solution of the system~(\ref{7a})-(\ref{7c}) reads
\begin{eqnarray}
\rho_1&=&-k^2 A \exp(i\theta)+ {\rm c.c.},
\label{8a}\\
u_1&=& -k \omega A \exp(i\theta)+ {\rm c.c.},
\label{8b}\\
\Phi_1&=& A \exp(i\theta)+ {\rm c.c.}.
\label{8c}
\end{eqnarray}
%
Here, the unknown complex function $A$ represents, at this 
order of approximation, the envelope of the perturbation of the background, which modulates
the carrier wave $\exp(i\theta)$; as we will show below, the function $A$ 
satisfies a spherical NLS equation.

Proceeding to the next order, $\mathcal{O}(\epsilon^2)$ (i.e., $j=2$), the system
(\ref{7a})-(\ref{7c}) becomes inhomogeneous, with the right-hand side terms given by:
\begin{eqnarray}
f_2 &=& - \frac{\partial \rho_{1}}{\partial T_1} 
-\frac{\partial u_1}{\partial R_1} 
- \frac{\partial}{\partial r}\left(\rho_1 u_{1}\right),
\label{9a}\\
g_2 &=& -\frac{\partial u_1}{\partial T_1} - \frac{\partial \rho_{1}}{\partial R_1}  
- \frac{\partial \Phi_{1}}{\partial R_1} 
-\rho_1 \frac{\partial u_{1}}{\partial t} - u_1 \frac{\partial u_{1}}{\partial r}  
-\rho_1 \frac{\partial \Phi_{1}}{\partial r}, 
\label{9b}\\
h_2 &=& -2\frac{\partial^2 \Phi_{1}}{\partial r \partial R_1}.
\label{9c}
\end{eqnarray}
The solvability condition for the system (\ref{7a})-(\ref{7c}) at this order (or equivalently for Eq.~(\ref{eqrho}) for $j=2$) is obtained by demanding that there be no terms proportional to $\exp(i \theta)$ on the r.h.s. of~\eqref{eqrho}, which in turn guarantees the absence of secular terms in the solution for $\rho_2$. In particular, the secular term is of the form 
\footnote{Notice that if we had only used the variables $R_1$ and $T_1$, the relevant term would have been 
of the form $ \exp(i \theta) \left(\partial_{T_1} -\omega' \partial_{R_1}\right)A$. This secular 
term vanishes as long as $A$ depends on $R=R_1-\omega'T_1 =R_1-VT_1$.}
$(V-\omega'(k))A_R\exp(i \theta)$,  
and its removal yields the following expression for the unknown velocity $V$:
\begin{equation}
V=\frac{k}{\omega} = \omega'(k).
\label{vg}
\end{equation}
This means that $V$ is actually equal to the group velocity 
$\omega'(k)\equiv \partial \omega/\partial k$, as derived from the Jeans dispersion 
relation~(\ref{ldrj}). 
In addition, at the same order, we obtain the following expressions for the 
unknown fields $\rho_2$, $u_2$, $\Phi_2$:
\begin{eqnarray}
\rho_2&=&-\frac{2}{3}k^4(2k^2-3) A^2\exp(2i\theta)+{\rm c.c.},
\label{10a}\\
\nonumber \\
u_2&=& -\frac{k^3(4k^4-7k^2+3)}{3\omega} A^2\exp(2i\theta)
+ \frac{i}{\omega} \frac{\partial A}{\partial R} \exp(i\theta)+{\rm c.c.},
\label{10b} \\
\nonumber \\
\Phi_2&=&\frac{1}{6}k^4(2k^2-3) A^2\exp(2i\theta)
+\frac{2i}{k} \frac{\partial A}{\partial R} \exp(i\theta) +{\rm c.c.}.
\label{10c}
\end{eqnarray}

Finally, at order $\mathcal{O}(\epsilon^3)$ (i.e., for $j=3$), the right-hand side terms
of the system (\ref{7a})-(\ref{7c}) are given by:
\begin{eqnarray}
f_3=&-&\frac{2}{VT_2} u_1 - \frac{\partial \rho_1}{\partial T_2} 
-\frac{\partial \rho_2}{\partial T_1} -\frac{\partial}{\partial R_1}(\rho_1 u_1)
-\frac{\partial}{\partial r}(\rho_1 u_2+\rho_2 u_1) 
-\frac{\partial u_2}{\partial R_1},  
\\ 
\nonumber \\
g_3=&-&\frac{\partial u_1}{\partial T_2} - \rho_1 \frac{\partial u_1}{\partial T_1}
-\frac{\partial u_2}{\partial T_1} - u_1 \frac{\partial u_1}{\partial R_1}
-\frac{\partial \rho_2}{\partial R_1} - \rho_1 \frac{\partial \Phi_1}{\partial R_1}
-\frac{\partial \Phi_2}{\partial R_1} 
\nonumber \\
&-&\rho_1 \frac{\partial u_2}{\partial t}
-\rho_1 u_1 \frac{\partial u_1}{\partial r} - \frac{\partial }{\partial r}(u_1 u_2) 
-\rho_1 \frac{\partial \Phi_2}{\partial r}-\rho_2 \frac{\partial \Phi_1}{\partial r},
\\ \nonumber \\
h_3=&-& \frac{2}{VT_2} \frac{\partial \Phi_1}{\partial r}-\frac{\partial^2 \Phi_1}{\partial R_1^2} 
- 2\frac{\partial^2 \Phi_2}{\partial r \partial R_1}.
\end{eqnarray}
At this order too, the solvability condition for Eqs.~(\ref{7a})-(\ref{7c}) (or equivalently for Eq.~(\ref{eqrho})), is also obtained by removing the secular terms, $\propto \exp(i \theta$).
This way, we derive the following NLS equation for the envelope function $A$:

\begin{equation}
i \frac{\partial A}{\partial T_2}+\frac{i}{T_2}A+\frac{1}{2}\omega''(k) 
\frac{\partial^2 A}{\partial R^2}+\gamma(k) |A|^2 A=0,
\label{NLS}
\end{equation}
where the dispersion and nonlinearity coefficients, $\omega''(k)$ and $\gamma(k)$, are given by:
\begin{eqnarray}
\omega''(k) &\equiv& \frac{\partial^2 \omega}{\partial k^2} =
-\frac{1}{\omega^{3}}, 
\label{omdis} \\
\nonumber \\
\gamma(k)&=&\frac{k^4(8k^4-21k^2+12)}{6 \omega}.
\label{gama}
\end{eqnarray}
The functional forms of $\omega''(k)$ and $\gamma(k)$ are depicted in Fig.~\ref{omgamo}, 
in the wavenumber regime $k>k_J$ (i.e., in the regime of interest, free of the Jeans instability). 
It is observed that the dispersion coefficient is $\omega''(k)<0~\forall k$, while 
the nonlinearity coefficient becomes zero for $k=\alpha k_J$, where the value of parameter 
$\alpha$ is given by $\alpha =(1/4)\sqrt{21+\sqrt{57}}$. For $k_J<k<\alpha k_J$
we have $\gamma(k)<0$, while for $k>\alpha k_J$ we have $\gamma(k)>0$.   

\begin{figure}[t]
\centerline{\includegraphics[scale=0.39]{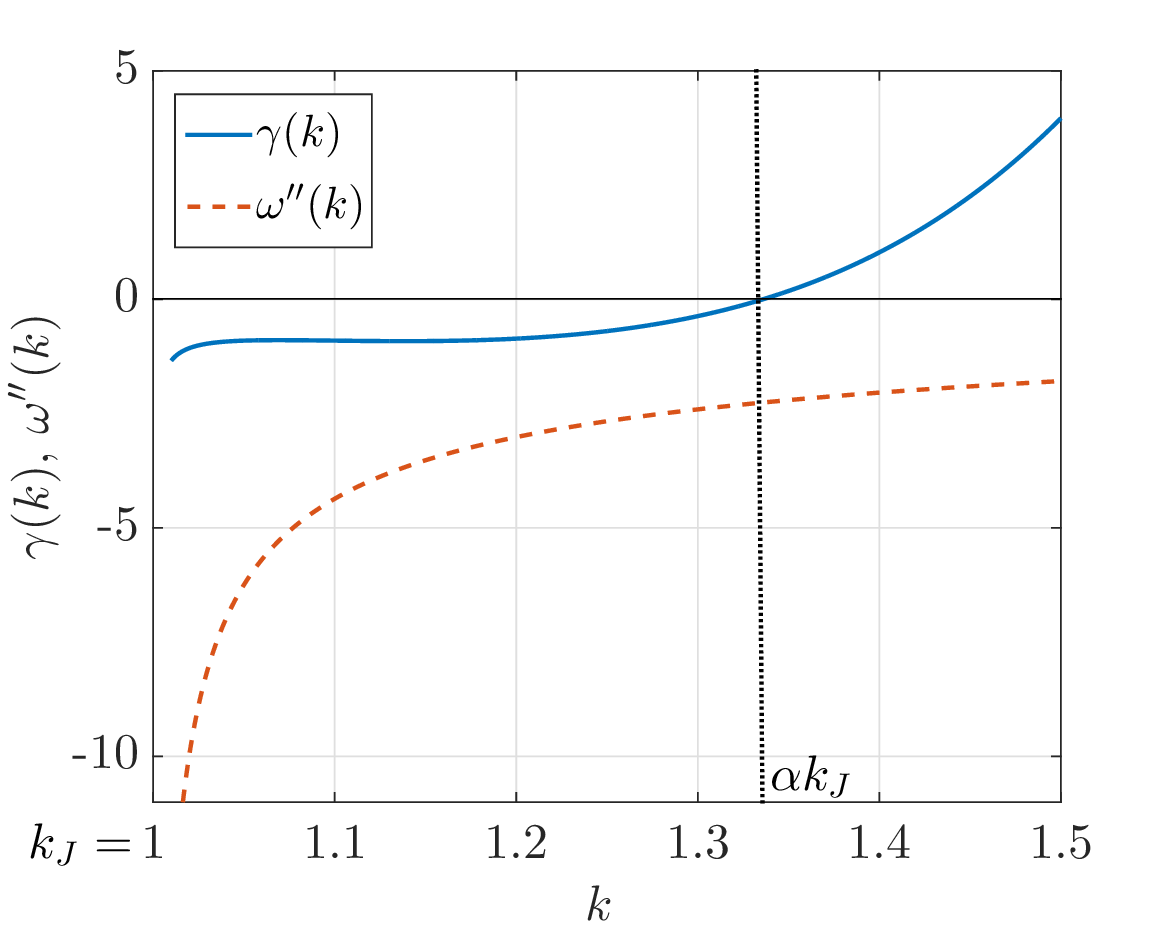}}
\caption{The dependence of the dispersion [solid (blue) line] and nonlinearity 
[dashed (red) line] coefficients, 
$\omega''$ and $\gamma$, on the wavenumber $k$, in the regime $k>k_J$. The dispersion 
coefficient is always negative, while the nonlinearity coefficient becomes zero at 
$k=\alpha k_J$, with $\alpha =(1/4)\sqrt{21+\sqrt{57}}$, and it is negative (positive) 
for $k<\alpha k_J$ ($k>\alpha k_J$).}
\label{omgamo}
\end{figure}

\section{The nonlinear instability}
\label{nMI}

We proceed with our analysis upon identifying an instability which stems from the NLS
Eq.~(\ref{NLS}). We will show that the latter possesses a spatially homogeneous 
and time-dependent solution, which is prone to modulational instability. This  
instability that occurs in the nonlinear regime of the system also depends on the 
curvature and, in some cases, may be arrested. 
This picture complements the one obtained in the linear regime of the problem, where 
solely the Jeans instability is present.

We start by introducing the transformation:
\begin{equation}
A(R,T_2) = \frac{T_0}{T_2}~\Psi(R,T_2),
\label{metasx}
\end{equation}
where $T_0=\epsilon^2 t_0$ is an initial time (see below). Then, by substituting in 
the NLS Eq.~(\ref{NLS}), it is straightforward to find that 
the function $\Psi$ satisfies the following equation:
\begin{equation}
i \frac{\partial \Psi}{\partial T_2} 
+\frac{1}{2}\omega''(k)  
\frac{\partial^2 \Psi}{\partial R^2}+\gamma(k)\frac{T_0^2}{T_2^2}~|\Psi|^2 \Psi=0. 
\label{1DNLSb}
\end{equation}
Observe that Eq.~(\ref{1DNLSb}) is actually a traditional $1$D NLS equation, but 
with the important difference that the nonlinearity coefficient $\gamma(k)$ is 
now replaced by $\gamma(k)(T_0/T_2)^2$. Hence, in other words, the spherical 
NLS equation~(\ref{NLS}) can be viewed as a traditional NLS, which 
gets linearized, becoming a linear Schr{\"o}dinger equation, in the asymptotic limit 
of $T_2 \rightarrow +\infty$. 

Next, we proceed by identifying an elementary, spatially homogeneous and 
time-dependent solution of the NLS Eq.~(\ref{1DNLSb}), which is of the form:
\begin{equation}
\Psi_b(T_2)=\Psi_0 \exp[i \Theta(T_2)],
\label{bg}
\end{equation}
where $\Psi_0$ is the constant amplitude of the solution, assumed to be real, and 
$\Theta(T_2)$ is a phase, with the associated nonlinear frequency shift, 
$d\Theta/dT_2$, satisfying the equation:
\begin{equation}
\frac{d \Theta}{d T_2} = \gamma(k)\Psi_0^2 \frac{T_0^2}{T_2^2}.
\label{dtheta}
\end{equation} 
Next, we wish to study the linear stability of the solution $\Psi_b(T_2)$. For this 
purpose, we introduce the linearization ansatz:
\begin{equation}
\Psi(R,T_2)=[\Psi_0+ (\delta \Psi)(R,T_2)]\exp[i \Theta(T_2)],
\end{equation}
where $|(\delta \Psi)|\ll \Psi_0$ is a small, generally complex, perturbation. 
Substituting into Eq.~(\ref{1DNLSb}), linearizing with respect to $\delta \Psi$, 
and also using Eq.~(\ref{dtheta}), we obtain:
\begin{equation}
i\frac{\partial (\delta \Psi)}{\partial T_2}
+\frac{1}{2}\omega''(k)  
\frac{\partial^2 (\delta \Psi)}{\partial R^2}+\gamma(k)\Psi_0^2\frac{T_0^2}{T_2^2} 
[(\delta \Psi) + (\bar{\delta \Psi})]=0,
\label{dpsi}
\end{equation}    
where $\bar{(\delta \Psi)}$ is the complex conjugate of $(\delta \Psi)$. Separating 
real and imaginary parts in $(\delta \Psi)$, namely $(\delta \Psi) = (\delta \Psi)_r 
+i (\delta \Psi)_i$, Eq.~(\ref{dpsi}) leads to the following system of linear PDEs: 
\begin{eqnarray}
&&-\frac{\partial (\delta \Psi)_i}{\partial T_2} 
+\frac{1}{2}\omega''(k)  
\frac{\partial^2 (\delta \Psi)_r}{\partial R^2}
+2\gamma(k)\Psi_0^2\frac{T_0^2}{T_2^2} (\delta \Psi)_r = 0,
\label{lp1} \\  
&&\frac{\partial (\delta \Psi)_r}{\partial T_2} 
+\frac{1}{2}\omega''(k)  
\frac{\partial^2 (\delta \Psi)_i}{\partial R^2}=0.
\label{lp2}
\end{eqnarray}
Then, the compatibility condition 
$\partial^3 (\delta \Psi)_i)/(\partial^2 R \partial T_2)
=\partial^3 (\delta \Psi)_i)/(\partial T_2 \partial^2 R )$ leads to the following
equation for $(\delta \Psi)_r$:
\begin{equation}
\frac{\partial^2 (\delta \Psi)_r}{\partial T_2^2} 
+\frac{1}{4}(\omega''(k))^2  
\frac{\partial^4 (\delta \Psi)_r}{\partial R^4}
+\gamma(k)\omega''(k)\Psi_0^2\frac{T_0^2}{T_2^2} 
\frac{\partial^2 (\delta \Psi)_r}{\partial R^2}= 0.
\label{tdB}
\end{equation}
To this end, considering solutions of the form,
\begin{equation}
(\delta \Psi)_r(R,T_2) = \psi_0(T_2) \exp(-iKR) +{\rm c.c.},
\end{equation}
we cast Eq.~(\ref{tdB}) into the form:
\begin{equation}
\frac{\partial^2 \psi_0}{\partial T_2^2} + \Omega^2(T_2) \psi_0=0,
\label{2ndovc}
\end{equation}
where 
%
%
\begin{equation}
\Omega^2(T_2) = \frac{1}{4}(\omega''(k))^2K^2 \left(K^2 -K_c^2\right),
\quad K_c^2(T_2) = \frac{4\gamma(k)A_0^2}{\omega''(k)}~\frac{T_0^2}{T_2^{2}},
\label{igr}
\end{equation}
with $A_0=\Psi_0$ (see Eq.~(\ref{metasx}) for $T_2=T_0$). 
Equation~(\ref{2ndovc}) is a 2nd-order ordinary differential equation (ODE) with a 
time-dependent coefficient, $\Omega^2(T_2)$. Obviously, if $\Omega^2(T_2)<0$ or, 
in other words, if
\begin{equation}
\gamma(k) \omega''(k) > 0, \quad {\rm and} \quad |K|<K_c,
\label{MIcond}
\end{equation}
then the elementary solution~(\ref{bg}) is prone to modulational instability (MI). 
Since $\omega''<0$ for every $k$ [see Eq.~(\ref{omdis}) and 
Fig.~\ref{omgamo}], 
the first of the above MI conditions is satisfied when $\gamma(k)<0$. As discussed 
above, the latter condition is fulfilled in the interval [see Eq.~(\ref{gama})]: 
\begin{equation}
k_J < |k| < \alpha k_J, \quad \alpha =\frac{1}{4}\sqrt{21+\sqrt{57}} \approx 1.34,
\label{extens}
\end{equation}
which defines the nonlinear instability band; this is highlighted by a red 
line-segment in Fig.~\ref{imoF}. Hence, our analysis reveals that the 
Jeans instability band, $|k|<k_J$, is extended by a factor $\alpha$ (i.e., up to 
$\approx 33.6\%$), when the dynamics of the full nonlinear problem is taken into 
account.

The dependence of the critical wavenumber $K_c$ on time renders  
the instability growth rate, Im$(\Omega)$, also a function of time (this is a 
typical situation occurring in the context of the spherical and cylindrical NLS 
equations, appearing in studies on plasma physics ---see, e.g., 
Refs.~\cite{shukla1,xue1,Demiray}). In particular, the growth rate Im$(\Omega)$, as     
can be found from Eq.~(\ref{igr}), is of the form:
\begin{equation}
{\rm Im}(\Omega)(K,T_2) = \frac{1}{2}|\omega''(k)| K \sqrt{K_c^2(T_2) - K^2},
\label{ngr}
\end{equation}
and is depicted in Fig.~\ref{imo2} as a function of $|K|<K_c$ for three different 
values of the slow time $T_2$, namely for $T_2=0.005$, $0.008$ and $0.01$, 
corresponding to the later stages of the evolution, 
e.g., for $t=0.5$, $0.8$, $1$ and $\epsilon=0.1$. (The carrier wavenumber has been fixed to $k=(1+\alpha)/2 \approx 1.17$, 
i.e., in the middle of the nonlinear instability band.)
It is important to point out that, as shown in Fig.~\ref{imo2}, the maximum growth rate of the 
nonlinear instability is of the same order of magnitude as the linear Jeans instability (see Fig.~\ref{imoF}). Furthermore, it is observed that, due to the dependence of $K_c$ on the slow time $T_2$, 
both the maximum growth rate and the range of the nonlinear instability band decrease 
with $T_2$. 
\begin{figure}[t]
\centerline{\includegraphics[scale=0.4]{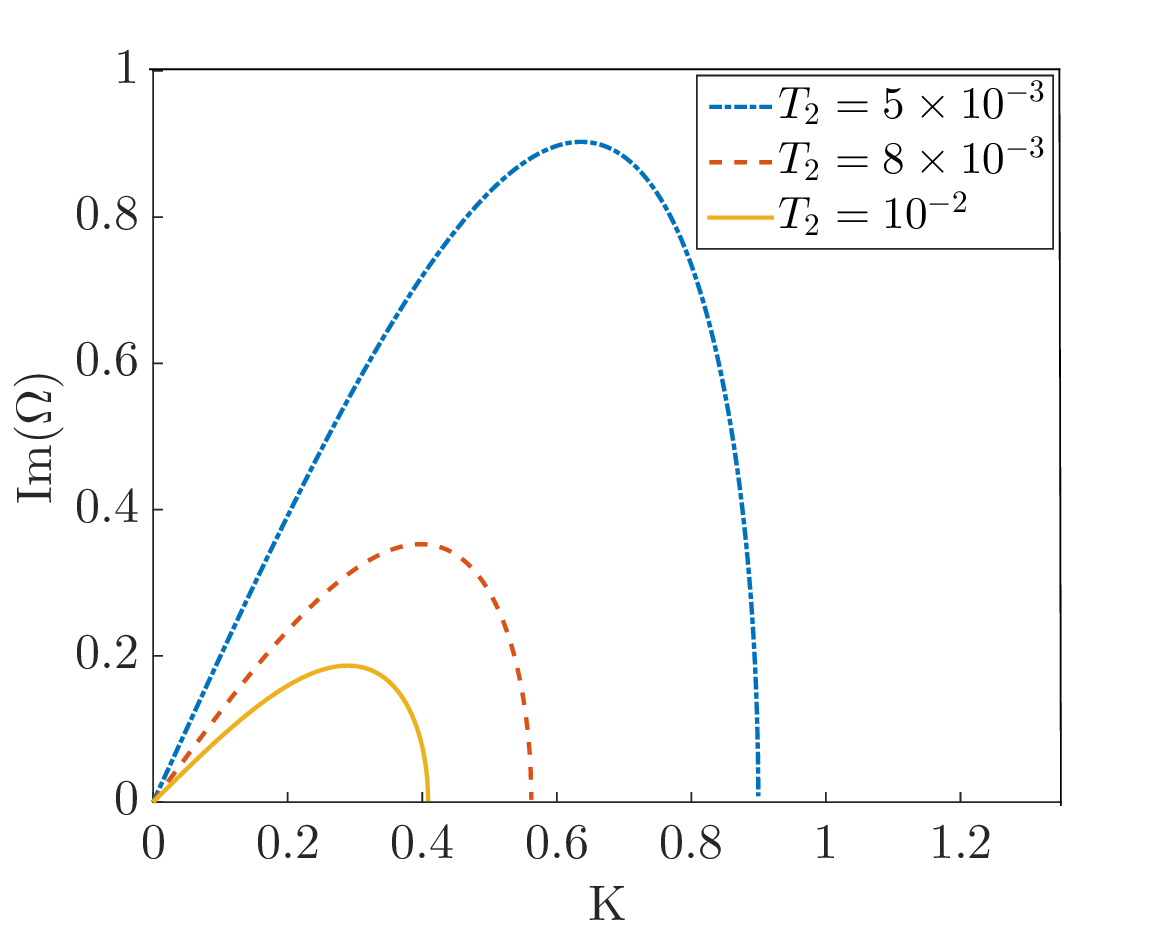}}
\caption{The instability growth rate, ${\rm Im}(\Omega)$ as a function of $K$ 
[see Eq.~(\ref{ngr})], 
for $k=(1+\alpha)k_J/2 \approx 1.17$ [i.e., in the middle of the nonlinearity-induced 
extension of the instability band ---see Eq.~(\ref{extens})], and for three different times:
$T_2=5\times 10^{-3}$ [upper, dashed-dotted (blue) curve], $T_2=8\times 10^{-3}$ 
[middle, dashed (red) curve], and $T_2=10^{-2}$ [lower, solid (orange) curve]. 
Notice that both the maximum value of ${\rm Im}(\Omega)$ and the range of the nonlinear 
instability band decrease with $T_2$. }
\label{imo2}
\end{figure}

An important remark at this point is the following. Contrary to the 1D case 
[when the curvature term $iA/T_2$ in Eq.~(\ref{NLS}) is absent], the nonlinear
instability ceases to exist for $K_c^2-K^2=0$, i.e., above some critical time 
$T_2^{\rm (c)}$ (a similar result was obtained in relevant studies in 
the context of plasma physics ---see, e.g., Refs.~\cite{shukla1,xue1,Demiray}). 
In particular, the instability growth will cease for $T_2\geq T_2^{\rm (c)}$, where, for a given wavenumber $K$, the 
critical time is given by: 
\begin{eqnarray}
T_2^{\rm (c)} &=& \frac{T_0 A_0}{K} \sqrt{\frac{4|\gamma(k)|}{|\omega''(k)|}}. 
\end{eqnarray}
Hence, the growth $\Gamma$ of the perturbation that takes place as long as the nonlinear instability occurs, i.e., from the initial time $T_0$ up to the time $T_2^{\rm (c)}$, is given by:
\begin{equation}
\Gamma(K) = \exp\left(\int_{T_0}^{T_2^{\rm (c)}} {\rm Im}(\Omega) dT_2'\right)
=\exp\left(A_{0}^2T_0 |\gamma(k)| F(w)\right) ,
\end{equation}
where the function $F(w)$ is defined as follows:
\begin{eqnarray}
F(w)= \frac{2}{w}\Bigg[\sqrt{w}\ln\left(\sqrt{w}+\sqrt{w-1}\right)
-\sqrt{w-1} \Bigg] , \quad \text{with}\quad w \equiv \dfrac{K_c^2(T_0)}{K^2} = \dfrac{4\gamma(k)A_0^2}{\omega''(k)K^2}.
\label{Fofw}
\end{eqnarray}
%
The function $F(w)$ is shown in Fig.~\ref{gr}. As seen therein, $F(w)$ 
initially increases, then reaches a maximum, namely
\begin{equation}
{\rm max} F(w_c)= \frac{2\sqrt{w_c-1}}{w_c}, 
\end{equation}
where $w_c$ is determined by the transcendental equation
\begin{equation}
2\sqrt{w_c-1}=\sqrt{w_c}\ln \left(\sqrt{w_c}+\sqrt{w_c-1}\right),
\end{equation}
while $F(w)$ decreases slowly thereafter, with $\lim_{w\rightarrow \infty}F(w)=0$. Note that
${\rm max} F(w_c) \approx 0.55$ and $w_c \approx 12$ (see Fig.~\ref{gr}).
Thus, according to the above discussion, while the growth of the MI is known to be 
always an increasing function of time in the 1D problem \cite{ZAKHAROV2009540}, this 
is not the case in the spherical geometry under consideration; hence, 
in the spherical geometry, wave patterns are expected to be, in principle, 
structurally more stable than the 1D ones. This can also be explained by the formal 
reduction of the spherical NLS [via the transformation of  Eq.~(\ref{metasx})] to the 
1D NLS~(\ref{1DNLSb}), with a time-dependent nonlinearity coefficient, 
$\propto 1/T_2^2$. This suggests that, at sufficiently large times, 
the NLS becomes a linear Schr{\"o}dinger equation as mentioned above, 
which in turn explains the fact that (a) the MI instability threshold is time-
dependent [see Eq.~(\ref{ngr})], and (b) wave structures of the spherical model are 
more stable than ones of the $1$D model.  

Let us now estimate the typical length scale of the instability as the wavenumber of the maximally growing mode. Since the only $K$-dependence of the growth factor comes via $F(w)$, which peaks at $w_{c}=12$, we have that  $K_c^2(T_0)/K_{\rm MI}^2=12$ and substituting $K_c(T_0)$ from~\eqref{igr}, we obtain (recall that $k = \epsilon K$)
\be
k_{\rm MI}(k)= \epsilon\sqrt{\frac{\gamma(k)}{3\omega''(k)}} A_0. 
\ee
%
For a carrier wave with $k=1.17$ the dispersion/nonlinearity coefficients read $|\omega''|=4.4$ and $|\gamma|=0.9$. Suppose that the background~\eqref{bg} becomes order one at some time $T_2=T_0A_0$. Then for $T_2$ close to the end of the dynamics, say $T_2 = 10^{-2}$, and $T_0 = 10^{-3}$, with $A_0\sim50$, we obtain a total growth of an order of magnitude, while the MI scale reads $k_{\rm MI} \approx 0.1$. Upon defining the characteristic length as $\lambda_{\rm MI} = 2\pi/ k_{\rm MI}$, we see that structures may emerge at scales larger than the Jeans length. This is compatible with the small-amplitude approximation that we have used, which holds true at the tails of galactic halos where the density is close to its equilibrium value. In relation to this let us also note that, while the function $F(w)$ clearly attains a maximum at $w_c\approx12$, it decreases quite slowly thereafter. This means that the instability persists for scales much larger than the Jeans length.
\begin{figure}[t]
\centerline{\includegraphics[scale=0.4]{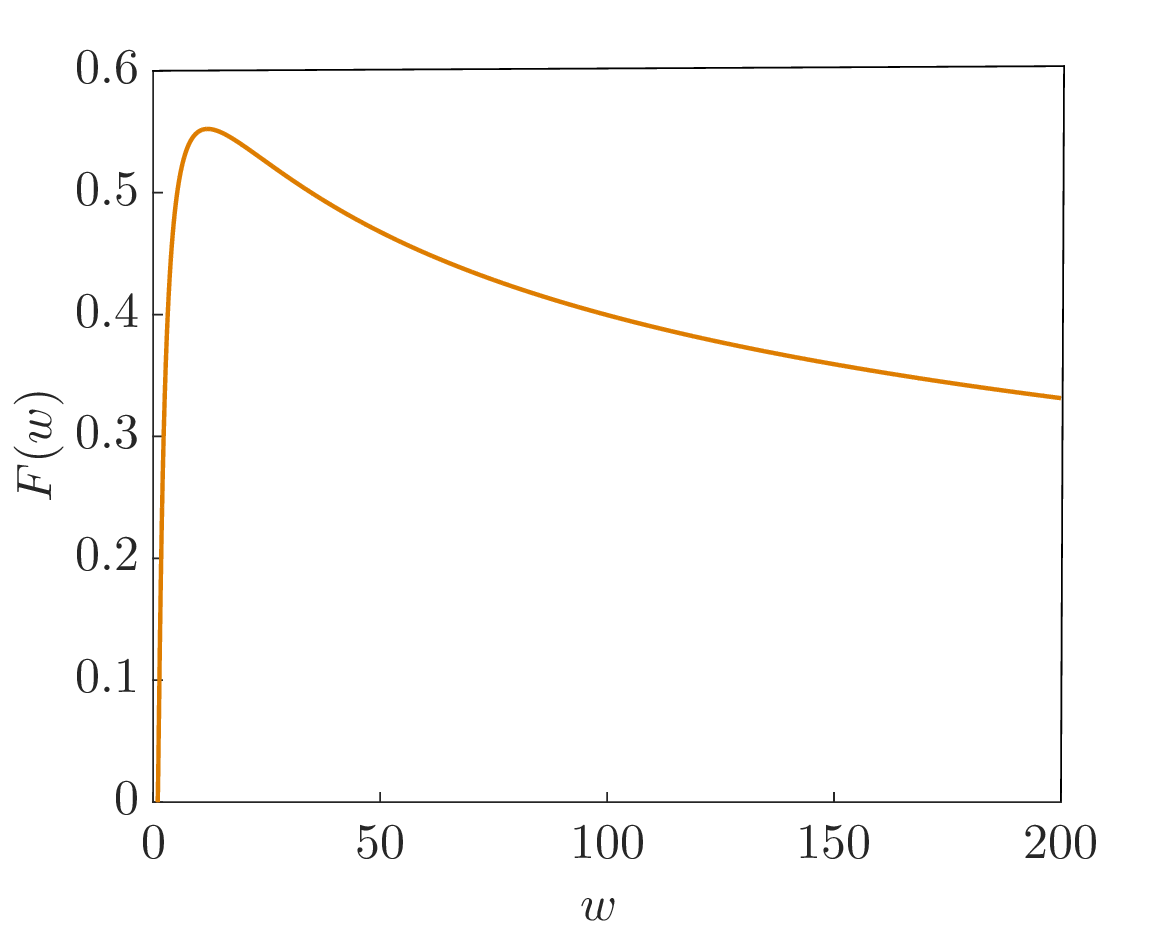}}
\caption{The function $F(w)$ given in Eq.~(\ref{Fofw}). It is observed that this function 
is initially increasing, then features a maximum ${\rm max} F(w_c) \approx 0.55$ at 
$w=w_c \approx 12$, and then it asymptotically decreases to zero.}
\label{gr}
\end{figure}

\section{Solitons and rogue waves}

The derivation of the effective NLS Eq.~(\ref{NLS}) that was presented above   
allows us to predict the existence of coherent wave structures, namely soliton solutions, 
which may describe large scale structures in the universe. 
Exact soliton solutions exist only in the 1D 
setting, i.e., in the absence of the curvature term $iA/T_2$. Nevertheless, as we will discuss 
below, approximate soliton solutions, analogous to the 1D case, are possible 
in the spherical geometry as well.   

Generally, soliton structures are supported by the NLS in the case where dispersion and 
nonlinearity are counterbalanced, i.e., when $|\omega''(k)/\gamma(k)| ={\cal O}(1)$.
Given the form of $\omega''(k)$ and $\gamma(k)$, this requirement is fulfilled
around the middle of the interval $k_J<k<\alpha k_J$ where the nonlinear instability occurs,  
as seen in Fig.~\ref{omgam}. Furthermore, the emergence of solitons is naturally 
connected with the initial conditions associated with the NLS equation. In that regard, it is 
necessary to make the following relevant comments. 

\begin{figure}[t]
\centerline{
\includegraphics[scale=0.39]{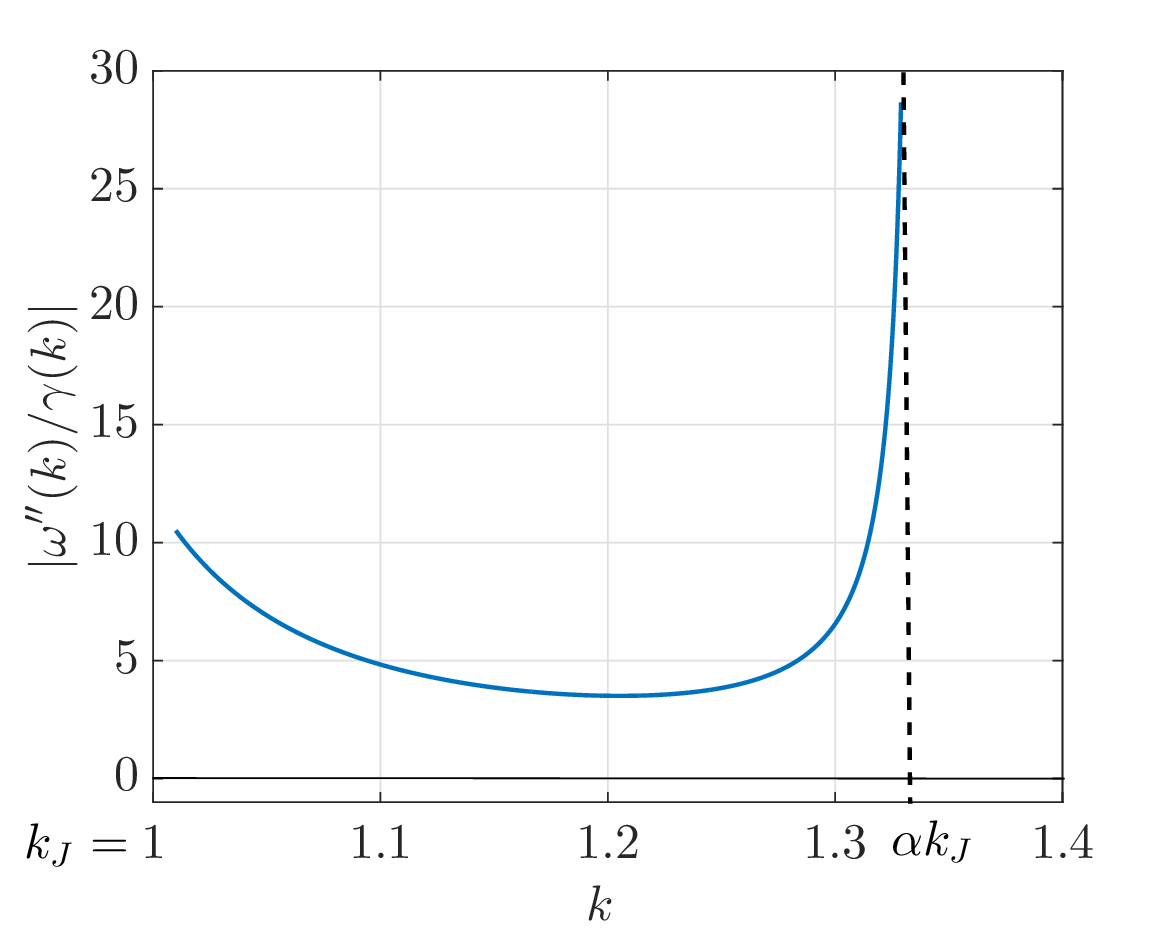}}
\caption{The ratio $|\omega''(k)/\gamma(k)|$ as a function of the wavenumber $k$ in the 
interval $k_J<k<\alpha k_J$, i.e., in the nonlinear instability band. As seen, this ratio 
is ${\cal O}(1)$ for wavenumbers around the middle of this interval.
}
\label{omgam}
\end{figure}

First, it is mentioned that coherent patterns can result from the onset of the 
nonlinear instability, i.e., the modulational instability (MI) mechanism discussed in 
the previous Section. Although MI was shown to occur for small-amplitude perturbations 
on top of the spatially uniform and time-dependent elementary solution~(\ref{bg}), any 
initial condition ---also in the form of a random perturbation (which is the 
physically relevant condition in our problem)--- can be viewed as a Fourier 
superposition of plane waves. Such a superposition will naturally contain 
wavenumbers that undergo MI 
\footnote{Here it should be stressed that we are referring to initial 
conditions characterized by wavenumbers which are Jeans stable; such a situation 
may occur in physically relevant parameter regimes, as explained in the end of 
Section~4.}
which, in turn, during its evolution, 
may give rise to localized coherent structures. Indeed, it can be shown (see, e.g., 
Ref.\cite{HasKod}) that at the early stage of its evolution, the MI results in 
the excitation of integer multiples of the perturbation wavenumbers of the unstable 
wave solutions, i.e., causes a widening of the wavenumber spectrum, which results 
in localization in space. This localization mechanism suggests the possibility of 
soliton formation. 
 
The most celebrated solutions of the NLS equation, which were found by means of the Inverse 
Scattering Transform method\cite{zakharov}, are the so-called ``bright solitons'' 
(this term is coined from the field of nonlinear optics\cite{kivsharagr}).
Such bell-shaped envelope solitons play a key role in the long-time dynamics of 
the NLS model, since they act as statistical attractors to which the system 
relaxes\cite{zturba,zturbb} (see also numerical evidence in Ref.\cite{turb}). 
Nevertheless, another important waveform 
that may be generated due to MI, when seeded from random perturbations of a background, 
is a single or a series of high-contrast peaks of random intensity; 
these peaks can be identified as 
``extreme events'' or ``rogue waves'' resulting from the unstable growth of weak wave 
modulations, which evolve into short groups of steep waves (see, e.g., the reviews\cite{onorato_report,tov} and references therein). These waves can be mathematically 
described by the so-called Peregrine soliton\cite{peregrine}. In contrast to 
the ``regular'' sech-shaped NLS soliton 
solution discussed above, the rogue wave is represented by a rational function, with 
the property of being localized in both time and space. Both of these important 
solutions will be discussed below.

We now proceed by presenting soliton solutions of the effective NLS Eq.~(\ref{NLS}).
First, in the purely 1D setting, the NLS equation is a completely integrable system, 
and possesses exact soliton solutions\cite{zakharov}. The NLS bright soliton is characterized by a pulse-shaped envelope
wave structure.
In our problem, it is relevant to consider the stationary form of the NLS soliton, namely:
\bea
A_{\rm sol}(R,T_2)&=&\sqrt{\frac{|\omega''|}{|\gamma|}}~\eta~{\rm sech}\left[\eta
\left(R-R_0\right)\right]\exp\left(i\eta^2|\omega''|T_2 \right),
\label{sol1d}
\eea
where $\eta$ is a free ${\cal O}(1)$ parameter, characterizing the amplitude, the inverse width and the frequency of the soliton, while $R_0$ represents the location of the soliton center. 

In addition, in the 1D setting, there exists another exact soliton solution, which is localized 
both in space and time and decays algebraically rather than exponentially, contrary to the soliton of Eq.~(\ref{sol1d}). The pertinent solution is the so-called Peregrine soliton\cite{peregrine} mentioned above, which is a rational solution of Eq.~(\ref{NLS}), of the following 
form:
\bea
A_{\rm RW}(R,T_2) &=& \sqrt{\frac{|\omega''|}{|\gamma|}} \eta
\left[1- \frac{4(1 + 2i \eta^2 |\omega''| T_2)}{1+ 4\eta^2 (R-R_0)^2 
+ 4 \eta^4 |\omega''|^2 T_2^2}\right]\exp(i\eta^2 |\omega''| T_2),
\label{RW}
\eea
where the free parameter $\eta$ is connected, as before, with the amplitude, 
inverse width and frequency of the rogue wave, while $R_0$ represents again the 
location of the center. Observe that the soliton waveform 
is on top of a finite background, decays to it asymptotically for either large $R$ or 
$T_2$ but exhibits a nontrivial behavior over a small region of $(R,T_2)$. That is 
why these localized solutions, that ``appear from 
nowhere and disappear without a trace''\cite{akhm} are considered as possible prototypes 
of rogue waves. 

Here, a couple of comments are in order. First, although the Peregrine soliton does not 
decay to zero as $R\rightarrow \infty$ (as was assumed in our multiscale perturbation scheme), 
the soliton core may well describe extreme wave events that may occur on top of the equilibrium 
density of the system (see relevant results and pertinent discussions in 
Refs.\cite{karan,horikis1,horikis2}).    
Second, in our problem, the time needed for the Peregrine soliton to disappear is very 
large (recall that $T_2=\epsilon^2 t$). Lastly, the solution~(\ref{RW}) suggests that fluctuations around the equilibrium density decay algebraically [rather than exponentially, 
as is the case corresponding to the soliton~(\ref{sol1d})].

Next, we proceed with the spherical geometry where, as mentioned above,
exact analytical soliton solutions of Eq.~(\ref{NLS}) are not available. Nevertheless,
approximate soliton solutions can be found in specific limiting cases, depending on the order
of magnitude of the terms involved in the NLS equation~(\ref{NLS}).
In particular, substituting either the soliton~(\ref{sol1d}) or the rogue
wave~(\ref{RW}) into the NLS and considering a specific time interval $\tau>0$
(i.e., $T_0 \le T_2 \le T_0+\tau$), we obtain:
\begin{eqnarray}
\frac{1}{T_2} A \propto \frac{\eta}{\tau}, \quad
\frac{1}{2}|\omega''(k)| A_{RR} \approx |\gamma(k)| |A|^2A \approx \eta^3,
\end{eqnarray}
in the aforementioned regime $|\omega''(k)/\gamma(k)| ={\cal O}(1)$, where 
1D solitons do exist. Then, two limiting cases can readily be identified:

\begin{itemize}
\item[(a)] The curvature effect is much stronger than the effects of dispersion
and nonlinearity, namely,
$$\frac{1}{T_2} A \gg \frac{1}{2}|\omega''(k)| A_{RR} \approx |\gamma(k)| |A|^2A,$$
\end{itemize}
for which case we have $\eta^2 \tau \ll 1$. In other words, 
%
%
in a situation where the initial amplitude $\eta$ is sufficiently small  
and the time interval $\tau$ is sufficiently short, 
the dispersive and nonlinear terms of Eq.~(\ref{NLS}) can be 
neglected. Then, in this limit, the NLS is reduced to the linear equation:
\begin{equation}
A_{T_2} + \frac{1}{T_2} A =0,
\label{AB}
\end{equation}
%
%
which, upon integration, leads to the following expression for $A(R, T_2)$:  
\begin{equation}
A(R, T_2)= \frac{T_0}{T_2} F(R),
\label{damp}
\end{equation}
where $F(R)$ is an arbitrary function of $R$. Then, we may construct an approximate 
soliton solution upon assuming that $F(R)$ has initially the profile of a planar 
soliton or a rogue wave [i.e., $F(R)$ has the functional form given by 
Eq.~(\ref{sol1d}) or Eq.~(\ref{RW}) for $T_2=T_0$]. In such a case, Eq.~(\ref{damp})
dictates that the relevant waveform decays in time, according to the law $T_0/T_2$.
For instance, the approximate functional form of the planar soliton reads:
\bea
A(R,T_2) &\approx& \eta \frac{T_0}{T_2} \sqrt{\frac{|\omega''|}{|\gamma|}}~  {\rm sech}\left[\eta \left( R-R_0 \right)\right]
\exp\left(i \eta^2 |\omega''|T_0 \right),
\label{sol1ds} 
\eea
while the form of the rogue wave can be found in a similar manner. 

\begin{itemize}
\item[(b)] The curvature effect is much weaker than the effects of dispersion
and nonlinearity, namely,
$$\frac{1}{T_2} A \ll \frac{1}{2}|\omega''(k)| A_{RR} \approx |\gamma(k)| |A|^2A,$$
\end{itemize}
for which case we have $\eta^2 \tau \gg 1$. Assuming now that $\eta={\cal O}(1)$, 
the latter inequality suggests that for sufficiently large times ($\tau \gg 1$), 
the effects of dispersion and nonlinearity dominate over the curvature effect. 
This means that the latter can be considered as a small perturbation, whose effect 
can be studied in the framework of the so-called adiabatic
approximation of the perturbation theory for solitons\cite{yubo}. 
According to this approach, which can be applied in the case of the planar 
soliton~(\ref{sol1d}), the functional form of the
soliton remains unchanged but its amplitude becomes time-dependent and evolves 
according to the law $\eta = \eta_0 (T_0/T_2)^{2}$; here, $\eta_0$ is a constant 
setting the solitary wave amplitude at the initial time $T_2=T_0$. Thus, an 
approximate spherical soliton solution, valid in the regime of long times 
($\tau \gg 1$), takes the form:
\begin{eqnarray}
A(R,T_2) \approx \eta_0 \left(\frac{T_0}{T_2} \right)^{2} \sqrt{\frac{|\omega''|}{|\gamma|}}~ 
{\rm sech}\left[\eta_0 \left(\frac{T_0}{T_2} \right)^{2}
\left( R-R_0 \right)\right] 
\exp\left\{i |\omega''|  \int_{T_0}^{T_2} \eta_0^2 \left(\frac{T_0}{T_2} \right)^{4} dT_2' 
\right\}.
%
\label{sol3d}
\end{eqnarray}

To conclude this section, it is found that, in both aforementioned asymptotic regimes 
(a) and (b), the solitons undergo a slow (algebraic) decay when the curvature effect 
is taken into account. 
It is also pointed out that the analytical estimates for the bright soliton decay presented 
above have also been discussed in the context of plasma physics; there,   
direct numerical simulations performed in the framework of Eq.~(\ref{NLS}), were found to be 
in good agreement with the analytical results\cite{shukla1,xue1}. Similar findings were 
also reported in Ref.\cite{sabry} for the case of rogue waves. Both solutions describe spherical shells of radius $R_0$ which decay exponentially/algebraically as $R\rightarrow \infty$; in the cosmological context, such structures may appear 
at the rims of galactic halos, where the density $\rho$ is close to the background density $\rho_0$, as per 
our considerations in the multiscale expansion method.

\section{Conclusions}

In this work we have studied the nonlinear dynamics of a self-gravitating fluid 
model in spherical geometry, which may find applications in cosmic structure 
formation. 

The 1D version of the model without gravity, being a purely nonlinear hyperbolic 
system, gives rise to shock waves, i.e., the development of discontinuities in finite 
time. When gravity is included, the model features ---in addition to nonlinearity--- 
dispersion. In particular, small-amplitude (linear) excitations 
$\propto \exp[i(kr-\omega t)]$ (where $\omega$ and $k$ represent the frequency and 
the wavenmber) on top of the steady-state background are characterized by a dispersion 
relation of the form $\omega^2(k) = k^2 -k_J^2$, where $k_J$ is the so-called Jeans 
wavenumber. Obviously, since $\partial^2 \omega(k)/\partial k^2 \neq 0$, the system 
features dispersion and nonlinearity simultaneously, a fact suggesting the possibility 
of soliton formation.

In the linear regime described above, the particular form of the dispersion relation 
indicates the presence of an important instability, namely, the Jeans instability, 
which manifests itself for wavenumbers $|k| < k_J$. The Jeans instability occurs when 
gravity dominates the internal pressure, resulting in gravitational collapse and the 
formation of large-scale structures in the universe.  

Based on the above, we have tried to address two important questions: (a) apart 
from the Jeans instability that is predicted in the linear regime of the model, 
is there any other instability that may occur in the full nonlinear version of 
the model, and (b) since we are dealing with a nonlinear dispersive system, can we 
predict and derive soliton solutions? Our analysis suggests a positive answer to 
both questions.

In particular, we have employed a multiscale expansion method and derived from the 
original spherically symmetric self-gravitating fluid model an effective spherical 
nonlinear Schr{\"o}dinger (NLS) equation for the envelope of small-amplitude 
perturbations around the equilibrium state of the system. This NLS model allowed us 
to predict an instability occurring in the nonlinear regime of the system, namely 
the modulational instability (MI) of the weak, spatially uniform, time-varying 
solution of the NLS (this solution is the simplest nontrivial perturbation satisfying 
the NLS on top of the equilibrium state of the fluid model). We have found that, while 
the Jeans instability occurs for $|k|<k_J$, the MI sets in for 
$k_J <|k| < \alpha k_J$, with $\alpha \approx 1.34$. This means that Jeans stable 
wavenumbers may become unstable in the aforementioned MI band. Since at the early 
stage of the evolution the MI gives rise to widening of the wavenuber spectrum 
---which, in turn, causes localization in real space--- soliton formation is possible 
even outside the Jeans instability domain. In other words, in the physical problem 
under consideration, nonlinearity (and, in particular, the ``nonlinear instability'' 
that manifests itself in the nonlinear regime), when competing with the gravity-induced dispersion, may give rise to large scale structures in the universe.

Indeed, the derived effective NLS equation allowed us to predict such structures, 
which may be of the form of solitons or rogue waves. These structures feature, 
respectively, an exponential or an algebraic decay in space, while both of them 
experience a slow curvature-induced polynomial decay in time. The functional form of 
the predicted soliton states could be relevant for the description of the edges of 
dark-matter halos.
     
Our analysis and results suggest a number of interesting research directions for 
future studies. First of all, it would be relevant to perform direct numerical 
simulations in the spherical self-gravitating fluid model to better clarify the 
various possibilities investigated in this work. In addition, it would also be 
interesting to consider the effect of viscosity in the self-gravitating fluid. 
Furthermore, a study of a self-gravitating two-fluid model, with each fluid 
representing the luminous and the dark matter, is particularly challenging. The 
above themes are currently under investigation and relevant results will be 
presented elsewhere.

\section*{Acknowledgments}

This work is dedicated to Professor Ioannis G. Stratis, a close friend and 
collaborator, with whom we had the chance to have numerous discussions about science, 
research and academic life during his great career. Constructive discussions with  T. A. Apostolatos are kindly acknowledged. 
The research work of G.N.K. was supported by the Hellenic Foundation for Research
and Innovation (HFRI) under the HFRI PhD Fellowship grant (Fellowship
Number: $5860$). O.E. and S.S. have been supported by Thailand
NSRF via PMU-B (grant numbers B01F650006 and
B37G660013).

\end{document}